\newcommand{\be}{\begin{equation}}
\newcommand{\ee}{\end{equation}}
\newcommand{\ba}{\begin{eqnarray}}
\newcommand{\ea}{\end{eqnarray}}

\newcommand{\bea}{\begin{eqnarray}}
\newcommand{\eea}{\end{eqnarray}}

\newif\ifdraft
\drafttrue
\newif\ifpreprint
\preprinttrue

\def\sandp#1.#2.#3{%
\left\langle\smash{#1}{\vphantom1}^{-}\right|{#2}%
\left|\smash{#3}{\vphantom1}^{+}\right\rangle}
\def\sandpp#1.#2.#3{%
\left\langle\smash{#1}{\vphantom1}^{+}\right|{#2}%
\left|\smash{#3}{\vphantom1}^{+}\right\rangle}
\def\sandmm#1.#2.#3{%
\left\langle\smash{#1}{\vphantom1}^{-}\right|{#2}%
\left|\smash{#3}{\vphantom1}^{-}\right\rangle}
\def\spab#1.#2.#3{\sandmm#1.#2.#3}
\def\spba#1.#2.#3{\sandpp#1.#2.#3}
\def\spaa#1.#2.#3.#4{\sandmp#1.{#2#3}.#4}
\def\spbb#1.#2.#3.#4{\sandpm#1.{#2#3}.#4}
\def\spa#1.#2{\langle#1\,#2\rangle}
\def\spb#1.#2{[#1\,#2]}
\def\spash#1.#2{\vphantom{\hat K}\spa{\smash{#1}}.{\smash{#2}}}
\def\spbsh#1.#2{\vphantom{\hat K}\spb{\smash{#1}}.{\smash{#2}}}
\def\lor#1.#2{\left(#1\,#2\right)}
\def\sand#1.#2.#3{%
\left\langle\smash{#1}{\vphantom1}^{-}\right|{#2}%
\left|\smash{#3}{\vphantom1}^{-}\right\rangle}
\def\sandpp#1.#2.#3{%
\left\langle\smash{#1}{\vphantom1}^{+}\right|{#2}%
\left|\smash{#3}{\vphantom1}^{+}\right\rangle}
\def\sandpm#1.#2.#3{%
\left\langle\smash{#1}{\vphantom1}^{+}\right|{#2}%
\left|\smash{#3}{\vphantom1}^{-}\right\rangle}
\def\sandmp#1.#2.#3{%
\left\langle\smash{#1}{\vphantom1}^{-}\right|{#2}%
\left|\smash{#3}{\vphantom1}^{+}\right\rangle}

\newbox\SlashedBox
\def\slashed#1{\setbox\SlashedBox=\hbox{#1}
\hbox to 0pt{\hbox to 1\wd\SlashedBox{\hfil/\hfil}\hss}#1}
\def\hboxtosizeof#1#2{\setbox\SlashedBox=\hbox{#1}
\hbox to 1\wd\SlashedBox{#2}}

\newbox\charbox
\newbox\slabox
\def\s#1{{      
        \setbox\charbox=\hbox{$#1$}
        \setbox\slabox=\hbox{$/$}
        \dimen\charbox=\ht\slabox
        \advance\dimen\charbox by -\dp\slabox
        \advance\dimen\charbox by -\ht\charbox
        \advance\dimen\charbox by \dp\charbox
        \divide\dimen\charbox by 2
        \raise-\dimen\charbox\hbox to \wd\charbox{\hss/\hss}
        \llap{$#1$}
}}

\def\eqn#1{eq.~(\ref{#1})}

\def\tree{{\rm tree}}

\def\sandp#1.#2.#3{%
\left\langle\smash{#1}{\vphantom1}^{+}\right|{#2}%
\left|\smash{#3}{\vphantom1}^{+}\right\rangle}

\def\Den#1#2 {\prod\limits_{k=#1}^{#2} \spa{k}.{(k+1)}}

\newcommand{\Bmp}[1]{\langle #1\rangle}

\newcommand{\At}{A^{\tree}}

\newbox\ourfigbox
\def\SizedFigureWithCaption#1#2#3{%
\setbox\ourfigbox
  \hbox{\hss\epsfxsize #1 \epsfbox{#2}\hss}
\hbox to \wd\ourfigbox{\vbox{\noindent\copy\ourfigbox\break
\vskip -6mm      \hbox to \wd\ourfigbox{\hss#3\hss}}}
}

\def\llongrightarrow{%
\relbar\mskip-0.5mu\joinrel\mskip-0.5mu\relbar
     \mskip-0.5mu\joinrel\longrightarrow}
\def\inlimit^#1{\buildrel#1\over\llongrightarrow}
\def\dash{\hbox{-\kern-.02em}}

\newcommand{\Kf}[1]{K^{\flat}_{#1}}

\newcommand{\Kfm}[1]{K^{\flat,-}_{#1}}

\documentclass{PoS}
\usepackage{epsfig}

\title{
\vspace*{-2cm}
\begin{flushright}
{\rm \small
SLAC-PUB-13124\\
UCLA/08/TEP/4}\\
\end{flushright}
\vspace*{2cm}
Constructing QCD one-loop amplitudes
}

\ShortTitle{Constructing QCD one-loop amplitudes}

\author{\speaker{Darren Forde}%
  \thanks{Research supported in part by the US Department of
 Energy under contracts DE--FG03--91ER40662 and
 DE--AC02--76SF00515.}\\
  {Stanford Linear Accelerator Center, Stanford University, Stanford, CA 94309, USA,\\
    \&\\
    Department of Physics and Astronomy, UCLA, Los Angeles, CA 90095--1547, USA.}\\
       E-mail: \email{dforde@slac.stanford.edu}}

\abstract{In the context of constructing one-loop amplitudes using a
 unitarity bootstrap approach we discuss a general systematic
 procedure for obtaining the coefficients of the scalar bubble and
 triangle integral functions of one-loop amplitudes. Coefficients are
 extracted after examining the behaviour of the cut integrand as the
 unconstrained parameters of a specifically chosen parametersiation
 of the cut loop momentum approach infinity. 
}

\FullConference{8th International Symposium on Radiative Corrections \\
                October 1-5, 2007\\
                Florence, Italy}

\begin{document}

Measurements of new physics at the forthcoming experimental program at
CERN's Large Hadron Collider (LHC) will require a precise
understanding of processes at next-to-leading order (NLO). This places
increased demands for the computation of new one-loop amplitudes. This
in turn has spurred recent developments towards improved calculational
techniques.

Direct calculations using Feynman diagrams are in general
inefficient. Developments of more efficient techniques have usually
centred around unitarity techniques~\cite{Neq4Oneloop}, where tree
amplitudes are effectively ``glued'' together to form loops. The most
straightforward application of this method, in which the cut loop
momentum is in $D=4$, allows for the computation of
``cut-constructible'' terms only, i.e. (poly)logarithmic containing
terms and any related constants. QCD amplitudes contain, in addition
to such terms, rational pieces which cannot be derived using such
cuts. These ``missing'' rational parts can be extracted using cut loop
momenta in $D=4-2\epsilon$~\cite{Bern:1995db}. The greater difficulty
of such calculations has restricted the application of this approach,
although recent developments~\cite{Anastasiou:2006jv,Giele:2008ve}
have provided new promise for this technique.

Recently the application of on-shell recursion
relations~\cite{Britto:2004ap} to obtaining the
``missing'' rational parts of one-loop
processes~\cite{Bern:2005hs} has provided an alternative
very promising solution to this problem. In combination with unitarity
methods an ``on-shell bootstrap'' approach provides an efficient
technique for computing complete one-loop QCD
amplitudes~\cite{Forde:2005hh}. Additionally
other new methods have also proved fruitful for calculating rational
terms~\cite{Xiao:2006vr}.

Such developments have again refocused attention on the optimisation
of the derivation of the cut-constructable pieces of the
amplitude. Deriving cut-constructible terms for any one-loop amplitude
reduces to the computation of coefficients of a set of scalar bubble,
scalar triangle and scalar box integral functions.  Box coefficients
may be found with very little work, directly from the quadruple cut of
the relevant box function~\cite{Britto:2004nc}. A unique box
coefficient contributes to each distinct quadruple cut. Unfortunately
triangle and bubble coefficients cannot be derived in quite so direct
a manner. Multiple scalar integral coefficients appear inside a
two-particle cut or triple cut. It is therefore necessary to
disentangle the relevant bubble or triangle coefficients from any
other coefficients sharing the same
cut~\cite{Neq4Oneloop,Giele:2008ve,Britto:2005ha,Ossola:2006us}.  The
large number of NLO processes of interest for the LHC suggests that a
completely automated computational procedure is highly desired.  To
this end we discuss, in this proceeding, a recently proposed
method~\cite{Forde:2007mi,Kilgore:2007qr} for the direct, efficient
and systematic extraction of bubble and triangle coefficients which is
well suited to automation.

\section{Triangle coefficients}
\label{eq:triple_cuts_and_scalar_triangles}

Following in the spirit of the box coefficient~\cite{Britto:2004nc} we
would like to apply a triple cut to extract a triangle
coefficient. Such a triple cut isolates a unique triangle coefficient
but also contains contributions from scalar box integrals which share
three of their four propagators with the triangle. The separation of
the coefficient of a particular scalar triangle integral from any box
coefficients can be effected by
\begin{eqnarray}
c_0=-\left[\textrm{Inf}_{t} A_1A_2A_3\right](t)\Big|_{t=0}.\label{eq:scalar_triangle_MI}
\end{eqnarray}

\begin{figure}[ht]
\begin{center}
\begin{minipage}{0.04\textwidth}
\begin{center}
a)
\end{center}
\end{minipage}%
\begin{minipage}{0.38\textwidth}
\begin{center}
\epsfxsize 2.0 truein\epsfbox{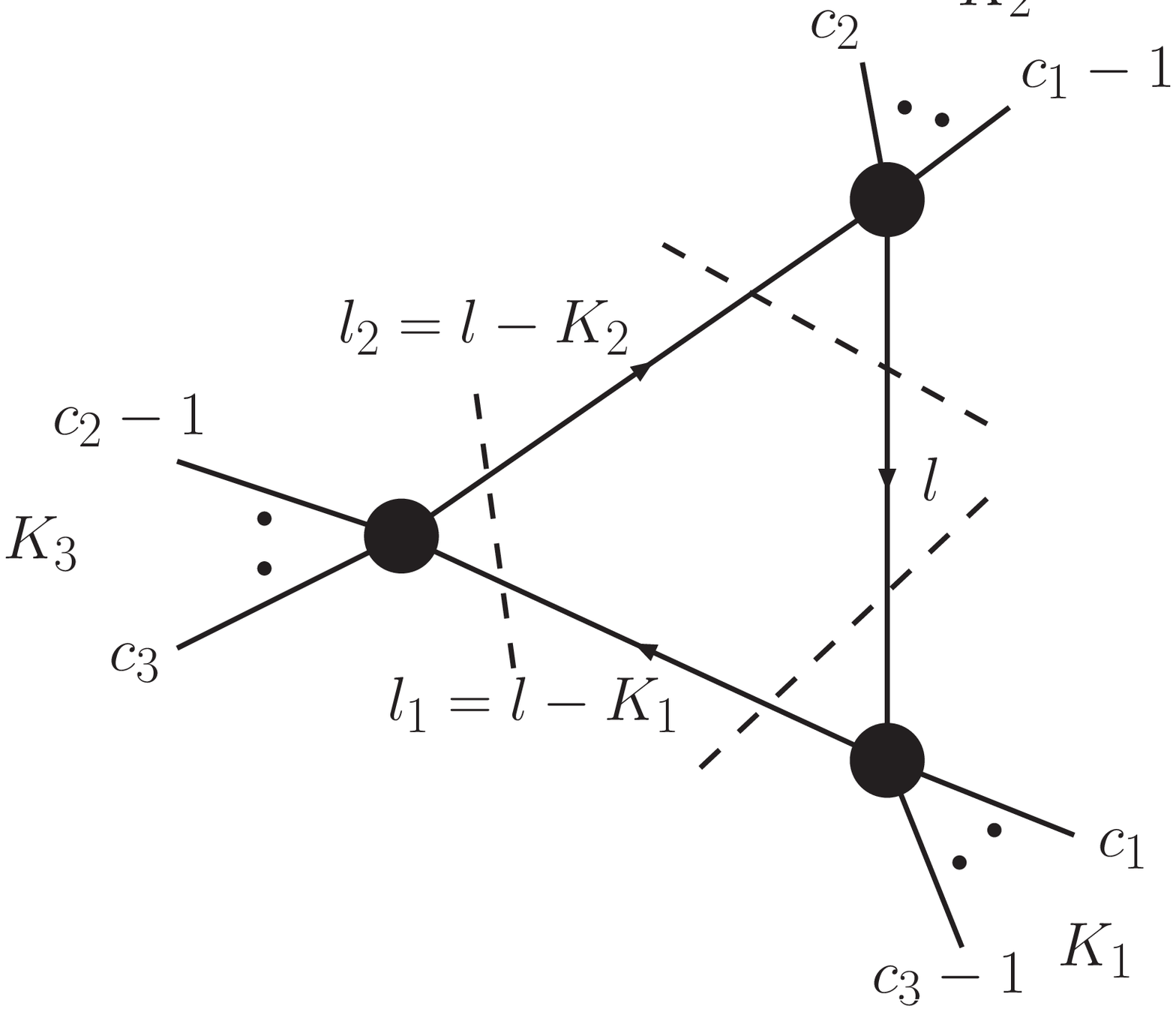}
\end{center}
\end{minipage}%
\begin{minipage}{0.04\textwidth}
\begin{center}
b)
\end{center}
\end{minipage}%
\begin{minipage}{0.42\textwidth}
\begin{center}
\epsfxsize 2.2 truein\epsfbox{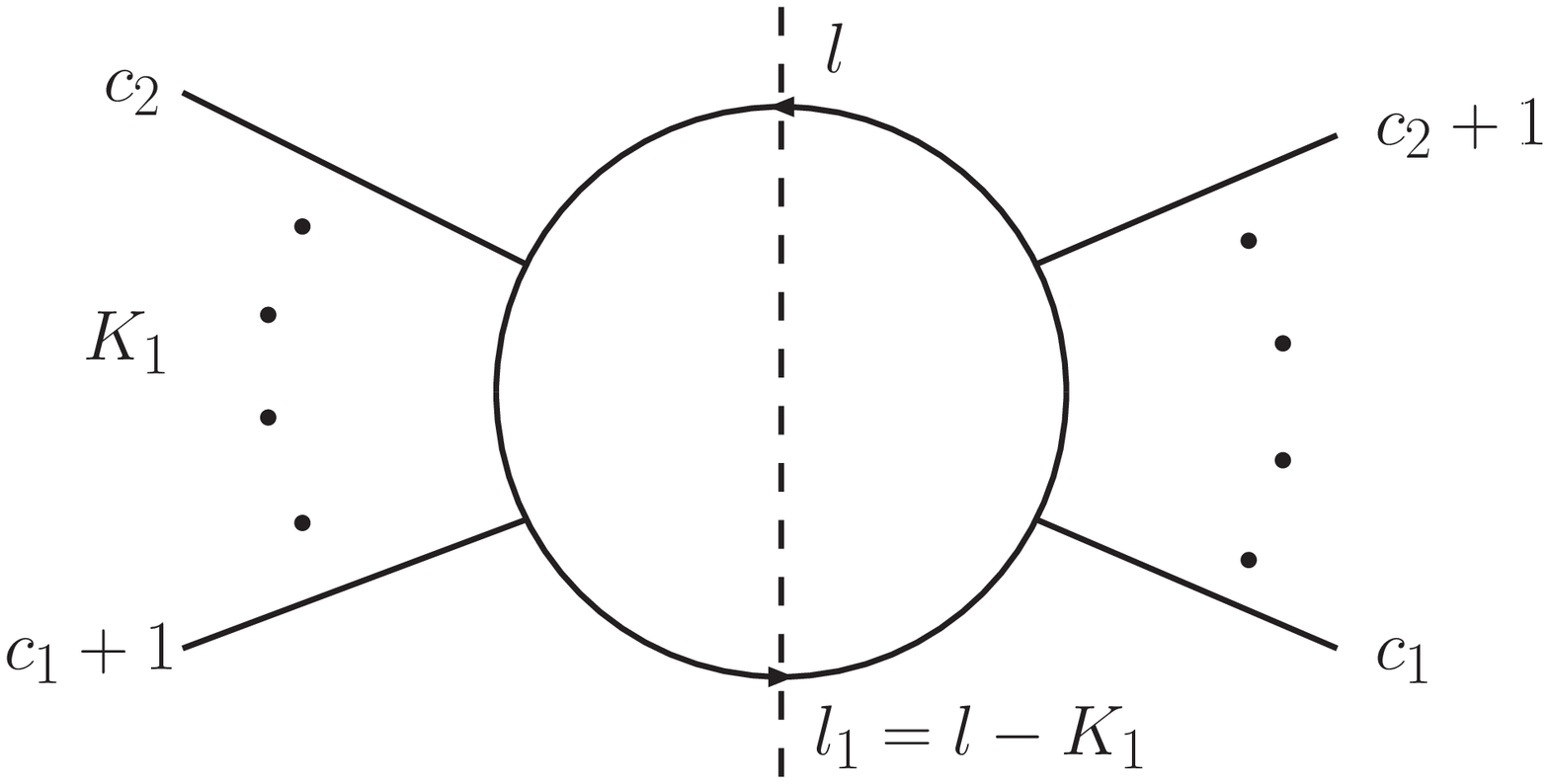}
\end{center}
\end{minipage}
\caption{ a) The triple cut used to compute a scalar triangle
coefficient. b) The two-particle cut used to calculate a scalar bubble
coefficient.  }
\label{figure:3_mass_triple_cut}
\end{center}
\end{figure}

Equation~\ref{eq:scalar_triangle_MI} instructs us to start by taking
the triple cut of the desired triangle coefficient,
\begin{eqnarray}
\At_1\At_2\At_3(t)&=&\At_{c_3-c_1+2}(-l,c_1,\ldots,(c_3-1),l_1) 
\At_{c_2-c_3+2}(-l_1,c_3,\ldots,(c_2-1),l_2)
\nonumber\\
&&\hphantom{\At_{c_3-c_1+2}(-l,c_1}\times\At_{n-c_2+c_1+2}(-l_2,c_2,\ldots,(c_1-1),l),
\end{eqnarray}
shown in figure~\ref{figure:3_mass_triple_cut}(a),
with $l_1=l-K_1$ and $l_2=l+K_2$, where $K_1$ and $K_2$ are sums of
external momenta. The cut momentum $l$ of the triple cut depends on a
single parameter $t$, and is parameterised in the specific
form~\cite{Ossola:2006us,Forde:2007mi},
\begin{eqnarray}
\langle l^-|=t\langle K_{1}^{\flat,-}|+\frac{S_1\left(\gamma-S_2\right)}{\left(\gamma^2-S_1S_2\right)}\langle K_{2}^{\flat,-}|,
&&\;\;\;\;\;
\langle
l^+|=\frac{S_2\left(\gamma-S_1\right)}{\left(\gamma^2-S_1S_2\right)t}\langle K_{1}^{\flat,+}|+\langle K_{2}^{\flat,+}|.
\label{eq:three_mass_mom_param}
\end{eqnarray}
Here we have expressed the cut momentum in terms of a convenient
basis of null vectors $\Kf1$ and $\Kf2$
\begin{eqnarray}
K_1^{\flat,\mu}=\frac{K^{\mu}_1-(S_1/\gamma)K^{\mu}_2}{1-(S_1S_2/\gamma^2)},
&&\;\;\;\;\;K_2^{\flat,\mu}=\frac{K^{\mu}_2-(S_2/\gamma)K^{\mu}_1}{1-(S_1S_2/\gamma^2)},\label{eq:tri_K1_K2_def}
\end{eqnarray}
with two solutions for $\gamma=(K_1\cdot K_2)\pm\sqrt{\Delta}$
with $\Delta=(K_1\cdot K_2)^2-K_1^2K_2^2$. 

The $\textrm{Inf}_{t}$ is instructing us to series expand this cut
integrand around $t=\infty$. The $t^0$ component of this series
expansion gives the desired triangle coefficient. For the three-mass
case described above we must also average over the two solutions to
$\gamma$. Analytic continuation of $l$ to complex momenta allows one-
and two-mass triangles, containing three-point vertices, to be
computed in a similar manner after setting the relevant masses in
\eqn{eq:three_mass_mom_param} and \eqn{eq:tri_K1_K2_def} to zero. In these cases only one
solution to $\gamma$ survives.

This procedure succeeds because of the specific momentum
parametrisation we have chosen. The series expansion of the
$\textrm{Inf}_{t}$ would in general give us rational coefficients
$a_i$ multiplying integrals over powers of $t$. Seen schematically
this is
\begin{eqnarray}
\sum_{i=-\infty}^{-1}a_i \int dt \;t^i+a_0\int dt+a_1 \int dt
\;t+\ldots+a_{\rm max}\int dt \;t^{\rm max},
\end{eqnarray}
and we would expect contributions to the scalar triangle coefficient
from every term. It is easy to show though that all integrals over $t$
will vanish, eliminating any such contributions. For example,
\begin{eqnarray}
\int dt t\sim\int d^4l\frac{\Bmp{\Kfm2|\s l|\Kfm1}}{l^2l_1^2l_2^2}\sim\Bmp{\Kfm2|\s K_1|\Kfm1}\mathcal{C}_1+\Bmp{\Kfm2|\s K_2|\Kfm1}\mathcal{C}_2=0,
\end{eqnarray}
with a similar result for other non-zero powers of $t$. The
$\mathcal{C}_i$ are Passarino-Veltman reduction coefficients.


\section{Bubble coefficients}
\label{section:double_cuts_and_bubble_coeffs}

A similar procedure applies to the extraction of bubble
coefficients. To compute the coefficient of a particular bubble
we use a two-particle cut and must disentangle our desired coefficient
from the scalar boxes and triangles which this will also contain. We
start from the generic two-particle cut
$\At_1\At_2(t,y)=\At_{c_2-c_1+2}(-l,(c_1+1),\ldots,c_2,l_1)\,\,
\At_{c_1-c_2+2}(-l_1,(c_2+1),\ldots,c_1,l)$ shown in
figure~\ref{figure:3_mass_triple_cut}(b), with $l_1=l-K_1$. Having
isolated a single bubble coefficient we parameterise the cut loop
momentum $l$, which now depends upon two free parameters $t$ and $y$,
using
\begin{eqnarray}
\langle l^-|=t\langle
K_{1}^{\flat,-}|+\frac{S_1}{\gamma}\left(1-y\right)\langle
\chi^{-}|,&&\;\;\;\;\;
\langle l^+|=\frac{y}{t}\langle
K_{1}^{\flat,+}|+\langle \chi^{+}|.\label{eq:two_cut_param}
\end{eqnarray}
This is expressed in terms of a basis of massless on-shell momenta
$\Kf1$ and $\chi$. $\chi$ is an arbitrary free vector, which the
final result is independent of, used to define
$K_1^{\flat,\mu}=K_1^{\mu}-\left(S_1/\gamma\right)\chi^{\mu}$ with
$\gamma=\Bmp{\chi^\pm|\s K_1|\chi^{\pm}}\equiv\Bmp{\chi^\pm|\s
{\Kf1}|\chi^{\pm}}$.

The equivalent expression to
\eqn{eq:scalar_triangle_MI} is then given by
\begin{eqnarray}
b_0\!=\!-i\left[\textrm{Inf}_t\left[\textrm{Inf}_y
A_1A_2\right](y)\right](t)\Big|_{t\rightarrow 0,\;y^m\rightarrow
\frac{1}{m+1}}-\frac{1}{2}\sum_{\mathcal{C}_{\rm tri}}
\left[\textrm{Inf}_t A_1A_2A_3\right](t)\Big|_{t^m\rightarrow T(m)}.\label{eq:scalar_bubble_MI}
\end{eqnarray}
The first term is the natural extension of the single
$\textrm{Inf}_{t}$ of \eqn{eq:scalar_triangle_MI} to the case of two
free parameters. In this double series expansion we expand around
$y=\infty$ and then $t=\infty$ and again drop terms proportional to
$t$, because the corresponding integrals disappear. Integrals over $y$ are
non-vanishing though and are related to the scalar bubble integral,
$B_0(K_1^2)$, via $\int dy y^m=B_0(K_1^2)/(m+1)$.

The naively unexpected second term of \eqn{eq:scalar_bubble_MI}
involves a sum over all the triangles $\mathcal{C}_{\rm tri}$ that
contain the original two-particle cut. Writing the two-particle cut
integrand schematically in the form
\begin{eqnarray}
a_0(t)+a_1(t)y+\ldots+a_{\rm max}(t)y^{\rm max}+\sum_{\mathcal{C}_{\rm tri}}\frac{A_L(y_i(t),t)A_R(y_i(t),t)}{\xi_i\left(y-y_i(t)\right)},
\end{eqnarray}
allows us to understand why contributions from triangles
arise. Solving $(l(y,t)-K_{2})^2=0$, the additional propagator present
in the triangles in the last term above, gives us $y_i(t)$. Inserting
this solution into the momentum parameterisation of $l$ given by
\eqn{eq:two_cut_param} leaves us with the momentum parametrisation of the
triangles that we wish to separate from the bubble. This
parametrisation differs importantly from
\eqn{eq:three_mass_mom_param} in that the integrals over $t$ do not
vanish, as can be seen for example with
\begin{eqnarray}
\int dt t\sim\int d^4l\frac{\Bmp{\chi^-|\s l|\Kfm1}}{l^2l_1^2l_2^2}\sim\Bmp{\chi^-|\s K_1|\Kfm1}\mathcal{C}_1+\Bmp{\chi^-|\s K_2|\Kfm1}\mathcal{C}_2\neq0.
\end{eqnarray}
The remaining contributions to the bubble coefficient are then found
by relating these non-vanishing integrals over $t$ to scalar bubbles
using
\begin{eqnarray}
\!\!\!\!\!\!T(m)=\!\left(\frac{S_1}{\gamma}\right)^m\frac{\Bmp{\chi^-|\s 
  K_2|\Kfm1}^m(K_1\cdot K_2)^{m-1}}{\Delta^{m}}\!\left(\sum_{l=1}^{m}\mathcal{C}_{ml}\frac{S_2^{l-1}}{(K_1\cdot K_2)^{l-1}}\right)\!B^{\rm cut}_0(K^2_1),\label{eq:the_triangle_t_ints}
\end{eqnarray}
and $T(0)=0$. The coefficients $\mathcal{C}_{ml}$ are given by
\begin{eqnarray}
&&\mathcal{C}_{11}=\frac{1}{2},\;\;\;\mathcal{C}_{21}=-\frac{3}{8},\;\;\;\mathcal{C}_{22}=-\frac{3}{8},
\;\;\;\mathcal{C}_{31}=-\frac{1}{12}\frac{\Delta}{(K_1\cdot K_2)^2}+\frac{5}{16},\;\;\;\mathcal{C}_{32}=\frac{5}{8},\;\;\;\mathcal{C}_{33}=\frac{5}{16}.\nonumber
\end{eqnarray}

\newpage
I would like to thank Zvi Bern, Lance Dixon, David Kosower and Carola
Berger for many helpful discussions.



\end{document}